\documentclass[article,nojss]{jss}

\usepackage{thumbpdf,lmodern}

\usepackage{framed}

\author{Francisco Palm\'i-Perales\\Universidad de Castilla-La Mancha
   \And Virgilio G\'omez-Rubio\\Universidad de Castilla-La Mancha
   \AND Miguel A. Martinez-Beneito \\Universitat de Val\`encia}
\Plainauthor{F. Palm\'i-Perales, V. G\'omez-Rubio, M. A. Martinez-Beneito}

\title{Bayesian Multivariate Spatial Models for Lattice Data with \pkg{INLA}}
\Plaintitle{Bayesian Multivariate Spatial Models for Lattice Data with INLA}
\Shorttitle{The \pkg{INLAMSM} Package}

\Abstract{
The \pkg{INLAMSM} package for the \proglang{R} programming language provides a
collection of multivariate spatial models for lattice data that can be used
with the \pkg{INLA} package for Bayesian inference.  The multivariate spatial
models implemented include different structures to model the spatial variation
of the variables and the between-variables variability.  In this way, fitting
multivariate spatial models becomes faster and easier.  The use of the
different models included in the package is illustrated using two different
datasets: the well-known North Carolina SIDS data and mortality by three causes
of death in Comunidad Valenciana (Spain).
}

\Plainkeywords{INLA, spatial statistics, multivariate modeling, R}

\Address{
  Francisco Palm\'i-Perales, Virgilio G\'omez-Rubio\\
  Departamento de Matem\'aticas\\
  Escuela T\'ecnica Superior de Ingenieros Industriales de Albacete\\
  Universidad de Castilla-La Mancha\\
  Avda. Espa\~na, s/n\\
  02002, Albacete, Spain\\
  E-mail: \email{Francisco.Palmi@uclm.es, Virgilio.Gomez@uclm.es}\\
  URL: \url{https://becarioprecario.github.io/} \\\\
  
  Miguel A. Martinez-Beneito\\
  Departamento de Estad\'istica i Investigaci\'o Operativa\\
  Universitat de Val\`encia\\
  Doctor Moliner, 50\\
  46100, Burjassot (Valencia), Spain\\
  E-mail: \email{miguel.a.martinez@uv.es}\\
  URL: \url{https://www.uv.es/~mamtnez/inicio_es.html}
}

\usepackage{Sweave}
\begin{document}

\section{Introduction}
\label{sec:intro}

The integrated nested Laplace approximation \citep[INLA,][]{INLA} provides an
alternative to traditional Markov chain Monte Carlo
\citep[MCMC,][]{Gilksetal:1996} for Bayesian inference. The INLA methodology
focuses on estimating the posterior marginals of the model parameters instead
of their joint posterior distribution. INLA is implemented in the \pkg{INLA}
package for the \proglang{R} programming language, that provides a simple way to
fit models via the \code{inla()} function, which works in a similar way as
other functions to fit regression models such as \code{glm()} or \code{gam()}.

The \pkg{INLA} package implements several likelihoods, priors and latent
effects that can be used to build models. It is also capable of fitting models
with several likelihoods, which can be useful for multivariate modeling.
However, multivariate spatial models are not included. The \pkg{INLAMSM}
package adds a number of multivariate latent effects that implement well-known
multivariate spatial models for areal data. By fitting these models with INLA,
instead of MCMC, computing times should be reduced.

Lattice data are made of  $I$ areas, usually related to some
administrative boundaries, where data are collected. We will assume
that values of $K$ variables are obtained from each area. In addition,
an adjacency structure is defined by considering that areas with a shared
boundary are neighbors. The analysis of this type of datasets is often
made by resorting to multivariate regression models.

This paper is organized as follows. The remainder of this introduction includes
a review of current software for multivariate spatial modeling.
Section~\ref{sec:models} gives a description of the different multivariate
spatial models implemented in the \pkg{INLAMSM} package.   Two detailed
applications of the \pkg{INLAMSM} package and the corresponding results can be
founded in Section~\ref{sec:examples}. Finally, a summary and discussion
are provided in Section~\ref{sec:discussion}.

\medskip
The \proglang{R} programming language provides a number of standalone packages for
multivariate spatial analysis.  In the specific case of analyzing spatial point
patterns, some \proglang{R} packages are available, such as \pkg{spatstat}
\citep{spatstat} and \pkg{spatialkernel} \citep{spatialkernel}. Package
\pkg{spatstat} is able to model multitype point patterns as well as handling a
good deal of the models and methods for the analysis of spatial and
spatio-temporal point patterns such as estimators of the space-time
inhomogeneous K-function and pair correlation function.  Package
\pkg{spatialkernel} performs edge-corrected kernel density estimation and
binary kernel regression estimation for multivariate spatial point patterns.



\medskip
Regarding geostatistical data, \pkg{gstat} \citep{GSTAT,gstatRpackage} provides
functions to fit both univariate and multivariate models using different
types of kriging. Package \pkg{spBayes} \citep{spBayes} is able to fit a wide
variety of Gaussian spatial process models for univariate as well as
multivariate point-referenced data using efficient MCMC algorithms.

\medskip
Finally, in the case of areal data, \proglang{R} package \pkg{CARBayes}
\citep{CARBayes2013} offers the possibility of fitting a wide class of CAR
models using MCMC methods. This package provides a number of univariate and
multivariate likelihoods, and it also includes a multivariate CAR model with an
inverse-Wishart and a CAR prior proposed by \citet{Lerouxetal:2000} to
estimate the variability between the different variables and the spatial
variation, respectively.

\medskip
In addition to these packages, the \proglang{BUGS} language
\citep{BUGS,WINBUGS} is a flexible framework for the implementation of
multivariate spatial models for lattice data. Package \pkg{R2WinBUGS}
\citep{R2WINBUGS} provides a set of functions to call the \pkg{WinBUGS}
software from \proglang{R}.  Recently, \proglang{Stan} \citep{Stan:2017}
develops a flexible language for Bayesian inference that could also be used to
fit (multivariate) spatial models \citep{Morrisetal:2019}. The \pkg{Rstan}
package \citep{Rstan} is a convenient interface between \proglang{R} and
\proglang{Stan} to fit models with ease.  Both \proglang{BUGS} and
\proglang{Stan} rely on MCMC algorithms for model fitting and inference.


\medskip
A computationally efficient alternative to software based on MCMC algorithms
is described in \cite{INLA}. This has been implemented in an \proglang{R} package
called \pkg{INLA}, which is often referred to as \pkg{R-INLA} to distinguish it
from the main INLA methodology. INLA advantages include an easy way of
implementing hierarchical models and short computing times to
fit spatial or spatio-temporal models
\citep{INLABayesianSpatial,INLAspatialreview}. This software is designed for
Bayesian inference on latent Gaussian Markov random field models which
include (generalized) linear mixed and spatial and spatio-temporal models.
\pkg{INLA} can deal with lattice data as well as geostatistical data by means of
an approximation to continuous spatial processes based on stochastic partial
differential equations \cite[SPDE,][]{SPDE}.  This approach can be used to fit
log-Gaussian Cox processes \citep{Simpsonetal:2016} for spatial and spatio-temporal point patterns as
well \citep{SPDE}.  Multivariate models can be fit by considering different
likelihoods with shared terms. See, for example, \citet{SPDE}, 
\citet{INLAVirgilioAdvanced} and \citet{INLAbook:2020}.

\medskip
Recently, some \proglang{R} packages have been developed on top of \pkg{R-INLA}.
For instance, \pkg{inlabru} \citep{inlabru} uses \pkg{INLA} 
to fit spatial density surfaces and estimating
abundance in a spatial point process, gridded and georeferenced context.

\medskip
As our review has shown, there are only a few packages that can fit
multivariate spatial models in a Bayesian framework. Furthermore, in the case
of analyzing multivariate areal data using Bayesian inference, there are
options based on MCMC and INLA. INLA is a widely used alternative to MCMC and
often provides shorter computing times when fitting spatial models.  However,
there are not a set of functions to fit multivariate spatial models for lattice
data using \pkg{INLA}. Developing these latent effects is the main goal of this
paper, which will enable users to include multivariate spatial effects in their
models to consider different ways of measuring spatial cross-correlation
\citep{SainCressie:2002}.

Some of the models implemented in the \pkg{INLAMSM} package have been described
by several authors \citep{GelfandVounatsou:2003,CarlinBanerjee:2003}.
Furthermore, other authors
\citep{Jinetal:2007,MartinezBeneito:2013,MacNab:2018} provide different reviews
of multivariate spatial models that describe the models implemented and many
others that could be added to the package in the future.  Furthermore,
\citet{bookMSM:2019} provide a recent review of multivariate spatial models in
the context of disease mapping. 

\section{Models}
\label{sec:models}

\pkg{INLAMSM} implements different functions which correspond to different
multivariate spatial latent effects. These differ in structure, complexity and
number of hyperparameters.  However, all of them are defined in a multivariate
spatial context in which $i = 1, \ldots, I$ represents the spatial areas and $k
= 1, \ldots, K$ is used to index the variables measured in region $i$.

Random effects can be represented using a matrix $\Theta$ with entries
$\theta_{ik},\ i=1,\ldots,I,\ k=1,\ldots,K$. Hence, the $k$-th column of
$\Theta$ represent the spatial random effects associated to variable $k$.

A particular application of these multivariate spatial models is disease
mapping, as described in the examples in Section~\ref{sec:examples}.  In this
context, the number of the observed cases of $k$-th disease in the $i$-th
spatial area, $Y_{ik}$, is modeled as a Poisson random variable:

$$
Y_{ik} \sim Po(E_{ik} \cdot R_{ik})
$$
\noindent
where $E_{ik}$ and $R_{ik}$ represent the expected cases and the relative risk
for the $i$-th spatial area and the $k$-th disease, respectively. Then, the
logarithm of the relative risk is the sum of two terms:

$$
\log(R_{ik})= a_k + \theta_{ik}
$$
\noindent
where $a_k$ is the intercept of the $k$-th disease and $\theta_{ik}$ is the term
that models the spatial variability. Note that other covariates can be 
included in the linear predictor on the right hand side of the previous
equation.

\medskip
Following \citet{MartinezBeneito:2013}, the variability of a multivariate
spatial latent effect can be divided in two terms: variability between the
values of the variables measured in the same area and the spatial variability
corresponding to the values of different areas for a particular variable. Both
sources of variability are modeled with their respective variance matrices, that have
different number of hyperparameters.

\medskip
Finally, before explaining any model, some useful specifications are defined.
Let $\Theta_{i\cdot}$ ($i=1, ...,I$) denote the $i$-th row of matrix
$\Theta$ and $\Theta_{\cdot j}$ ($j=1, ...,J$) denote the $j$th column of 
matrix $\Theta$. In addition, on a matrix $A=[A_{\cdot 1}: \cdots : A_{\cdot
J}] $, operator $vec(\cdot)$ is defined as

$$
vec(A) = (A_{\cdot 1}^{\top}, \ldots, A_{\cdot J}^{\top})^{\top} .
$$

Note that all latent effects will be defined using the \code{rgeneric} latent
effect, which defines the latent effect using the representation of the Gaussian Markov random
field. This is described in the \pkg{INLA} documentation, that can be accessed
with \code{inla.doc("rgeneric")}, and also in Chapter 11 of
\citet{INLAbook:2020}.  This representation is essentially a multivariate
Gaussian vector with a sparse precision matrix.  Hence, $vec(\Theta)$ will be
defined using a Normal distribution with zero mean and a highly structured
variance matrix $\Sigma$, i.e.,

$$ 
vec(\Theta) \sim N(0, \Sigma) .
$$ 

Note that \pkg{INLA} works with the precision and that this is why
$\Sigma^{-1}$ will be required instead of the variance matrix $\Sigma$.
Furthermore, $\Sigma$ will be created so that it accounts for the association
between the different variables within areas as well as association between
areas for the same variable or disease, as explained above.  The new latent
effects included in the \pkg{INLAMSM} can be used as a guidance to develop
other types of multivariate spatial latent effects.


\subsection{Independent intrinsic MCAR}

One of the easiest options to model the variability within- and
between-variables is using an intrinsic CAR distribution \citep{Besagetal:1991}
for the spatial structure and a diagonal matrix for the covariance matrix
between variables, respectively. Specifically, $\Theta$ is modeled as 

$$
vec(\Theta) \sim N\left(0, \Lambda^{-1} \otimes (D-W)^{-1}\right)
$$
\noindent
where $D=diag(n_1,...,n_I)$ is a diagonal matrix with values $n_i$  (the
number of neighbors of region $i$) and $W$ is an adjacency matrix with entries
$W_{ij}$. Each entry $W_{ij}$ is equal to 1 if units $i$ and $j$ are neighbors
and 0 otherwise.

Let $\tau_k$ be the marginal precision of the $k$-th variable. When variables
are independent of each other, the between-variables precision matrix $\Lambda$
is a $k \times k$ diagonal matrix defined as 

\[
  \Lambda=
  \left[ {\begin{array}{cccccc}
   \tau_1 &   0    & \cdots & \cdots &    0   \\
   0      & \tau_2 &   0    &        & \vdots \\
   \vdots &   0    & \ddots & \ddots & \vdots \\
   \vdots &        & \ddots & \ddots &    0   \\
   0      & \cdots & \cdots &    0   &  \tau_K\\
  \end{array} } \right]
.
\]

Following \citet{Jinetal:2007}, matrix $\Lambda$ can be regarded as a
(precision) matrix for the association between variables within the same area
and (precision) matrix $(D-W)^{-1}$ to account for the association
between areas for a given variable.  In this case it is assumed that the
different variables are independent and this is why $\Lambda$ is a diagonal
matrix, but in other latent random effects introduced below $\Lambda$ will be a
dense matrix to model association between the different variables.

Thus, this model has $K$ hyperparameters, $ \{\tau_k \}^{K}_{k=1}$, equal to
the total number of variables. This model is computationally fast
because the number of hyperparameters is the lowest among the multivariate
spatial models implemented in the \pkg{INLAMSM} package, as it will be seen
below.  Therefore, this model can be used  as a baseline to compare a
na\"ive model with independent spatial terms with more complex
alternatives.

Hyperparameters are internally represented in \pkg{INLA} using the log-scale
so that they are not bounded, and the vector
of hyperparameters is $(\log(\tau_1), \ldots, \log(\tau_K))$.  Prior
distributions are assigned to the standard deviations instead of the precisions
\citep{Gelman:2006}. In particular, a uniform improper distribution between $0$
and $+\infty$ is assigned to each standard deviation:

$$
\sigma_k = \frac{1}{\sqrt{\tau_k}} \sim Un(0, +\infty ); \,\,\,\,\,\, k=1, \ldots, K.
$$

Note that \pkg{INLA} will report the results in the internal scale but we have
included several functions in the package to transform from the log-precision
into some other more meaningful scale. See Section~\ref{sec:examples} for
details. Furthermore, \pkg{INLA} provides a set of functions that can be
used to make this transformation \citep[see, for example,][Chapter 2]{INLAbook:2020}.



Finally, given that this is an improper distribution, it will be
necessary to add further constraints when fitting the model, similarly to the
one dimensional intrinsic CAR distribution \citep{BesagKooperberg:1995}. In this
case, a sum-to-zero constraint will be added to the random effects associated
to each variable.  This implies that a variable-specific intercept should be
added to the model as well.

\subsection{Independent proper MCAR}

A proper CAR distribution can be used to model the within-variables variability instead
of the intrinsic version. Here, a common spatial autocorrelation parameter
$\alpha$ for all the variables is introduced. Thus $\Theta$ is modeled as 

$$
vec(\Theta) \sim N\left(0, \Lambda^{-1} \otimes (D- \alpha \cdot W)^{-1}\right)
$$
\noindent
where matrices $D$ and $W$ are defined similarly as above.  The matrix used to
model the between-variables variability keeps the same structure, thus
$\Lambda$ is defined as a diagonal matrix with entries $(\tau_1, \ldots,
\tau_K)$.

The proper CAR specification described above is not implemented in the
\pkg{INLA} package.  In our opinion this is an additional contribution of our
\pkg{INLAMSM} package. It is worth noting that the way in which the proper CAR
distribution is implemented here differs from the proper latent random effects
implemented in the \pkg{INLA} package as defined in latent effects
\code{besagproper} and \code{besagproper2}. The \code{besagproper} effect is
based on adding a constant to the diagonal entries of the precision matrix of
the intrinsic CAR distribution, while the \code{besagproper2} effect is
actually the model proposed by \cite{Lerouxetal:2000}. 

Now, this model has $K+1$ hyperparameters which are the spatial autocorrelation
parameter, $\alpha$, and the $K$ precisions, $ \{\tau_k \}^{K}_{k=1}$.
Internally, the hyperparameters are $(\alpha^{*},\log(\tau_1), \ldots,
\log(\tau_K)).$ Hyperparameter $\alpha^{*}$ is defined by transforming $\alpha$
as follows:

$$
\alpha^{*} = logit(\frac{\alpha - \alpha_{min}}{\alpha_{max} - \alpha_{min}}) 
$$
\noindent
where $\alpha_{min}$ and $\alpha_{max}$ define the bounds of the domain of
$\alpha$ \citep{Sunetal:1999}. However, the proper CAR distribution has been
aduced to show weird artifacts when the spatial correlation parameter takes 
negative values \citep{Wall:2004}. For this reason, it is common to assume 
$\alpha_{min}=0$ and $\alpha_{max}=1$.

As in the previous model, uniform prior distributions are set on the standard
deviations, and a uniform prior on $\alpha$ is considered:

$$
\sigma_k =
\frac{1}{\sqrt{\tau_k}} \sim Un(0, +\infty); \,\,\,\,\,\, k=1, \ldots, K ,
$$

$$
\alpha \sim Un(\alpha_{min}, \alpha_{max}) .
$$ 

Once again, the independent proper MCAR distribution is a na\"ive
implementation of independent spatial patterns as the independent intrinsic
MCAR. Nevertheless, this model could also be used as a benchmark for comparing
multivariate proper CAR models assuming dependence between variables (as
described below).

\subsection{Improper MCAR model}

A diagonal precision matrix $\Lambda^{-1}$ is a na\"ive way to model the
between-variables variability which leads to short computing times. However,
setting the off-diagonal elements of $\Lambda$ to zero assumes that variables
are independent of each other.  The off-diagonal elements model dependence
between every pair of variables, which could be useful when looking for similar
spatial patterns of different variables. 

Therefore, a dense precision matrix and an intrinsic conditional autoregressive
distribution can be chosen to model the between- and within-variables
variability, respectively. This can be regarded as a generalization of the
univariate intrinsic conditional autoregressive model. In particular, $\Theta$
is modeled similarly as the independent IMCAR:

$$
vec(\Theta) \sim N\left(0, \Lambda^{-1} \otimes (D-W)^{-1}\right)
$$
\noindent
but now $\Lambda$ is a dense, symmetric and positive-definite matrix. The
parameterization of $\Lambda$ follows that of other latent effects implemented
in \pkg{INLA} and the hyperparameters are the precisions of the variables and
the correlations between any pair of variables. Hence, instead of dealing
with the structure of $\Lambda$, $\Lambda^{-1}$ is defined as follows:


\[
  \Lambda^{-1}= 
  \left[ {\begin{array}{cccccc}
             1/\tau_1          & \rho_{12}/\sqrt{\tau_1\tau_2} &           \cdots              &               \cdots                  &      \rho_{1K}/\sqrt{\tau_1\tau_K}    \\
 \rho_{12}/\sqrt{\tau_1\tau_2} &            1/\tau_2           & \rho_{23}/\sqrt{\tau_2\tau_3} &                                       &                  \vdots               \\
              \vdots           & \rho_{23}/\sqrt{\tau_2\tau_3} &           \ddots              &               \ddots                  &                  \vdots               \\
              \vdots           &                               &           \ddots              &               \ddots                  & \rho_{(K-1)K}/\sqrt{\tau_{K-1}\tau_K} \\
 \rho_{1K}/\sqrt{\tau_1\tau_K} &            \cdots             &           \cdots              & \rho_{(K-1)K}/\sqrt{\tau_{K-1}\tau_K} &                 1/\tau_K              \\
  \end{array} } \right]
\]

\noindent

Here, $\rho_{jk}$ is the correlation coefficient between variables $j$ and $k$
and $\tau_k$ is the marginal precision of the $k$-th variable. Therefore, the
set of hyperparameters contains $K$ marginal precisions and $K*(K-1)/2$
correlation parameters. In this case, a joint prior distribution is
considered for $\Lambda^{-1}$ instead of setting a prior distribution
for each hyperparameter. Thus, $\Lambda^{-1}$ follows a Wishart distribution as

$$\Lambda^{-1} \sim Wishart_K(r, R^{-1})
$$
\noindent
where $r$ is the number of degrees of freedom and $R^{-1}$ is a fixed symmetric
positive definite matrix of size $K \times K$. In our implementation, $r$ is
equal to $K$ \citep[following ][]{CarlinBanerjee:2003} and $R$ is the $K\times
K$ identity matrix.

The vector of hyperparameters contains now the precisions plus the correlations
in the lower triangular matrix of $\Lambda^{-1}$ (columnwise). However, these
parameters are re-scaled so that the log-precisions are used and the correlation
parameters are transformed using

$$
\rho^{*} = logit((\rho + 1) / 2),
$$
\noindent
for any correlation $\rho$. Hence, the vector of hyperparameters in the
internal scale is now defined as $(\log(\tau_1), \ldots, \log(\tau_K), \rho^{*}_{21}, \rho^{*}_{31}, \ldots)$.

As stated above, given that this distribution is improper, a sum-to-zero
constraint on the effects for each variable will be added when fitting the
model. This also means that variable-specific intercepts are required in the
model.

\subsection{Proper MCAR model}

A proper CAR distribution can be used instead of the intrinsic CAR in order to
model the within-variables variability. This alternative corresponds to the
multivariate generalization of the univariate proper conditional autoregressive
model. As in the previous case, a dense $\Lambda$ matrix is used to model the
between-variables variability. Specifically, $\Theta$ is modeled as 

$$
vec(\Theta) \sim N\left(0, \Lambda^{-1} \otimes (D-\alpha \cdot W)^{-1}\right)
$$
\noindent
where $\alpha$ is a common spatial autocorrelation parameter and $\Lambda$ is a
dense symmetric positive-definite matrix with $\Lambda^{-1}$ defined as in the previous model.

In this case, the set of hyperparameters comprehends one common spatial
autocorrelation parameter $\alpha$, $K$ marginal precisions, $\{\tau_k
\}^{K}_{k=1}$, and $K*(K-1)/2$ correlation parameters, $
\{\rho_{jk}\}^{K}_{j,k=1}$.  Again, a Wishart distribution is
considered as a joint prior distribution for the $\Lambda^{-1}$ matrix, while a
uniform prior is consider for $\alpha$, i.e., 

$$
\alpha \sim Un(\alpha_{min}, \alpha_{max}) .
$$ 

In this case, the vector of hyperparameters contains the spatial
autocorrelation parameter $\alpha$, the precisions and the correlations in the
lower triangular elements of $\Lambda^{-1}$ (columnwise). As in previous
models, all hyperparameters are transformed to take values in the $(-\infty,
+\infty)$ interval.  Hence, the vector of hyperparameters is $(\alpha^{*},
\log(\tau_1),\ldots,\log(\tau_K), \rho^{*}_{21}, \rho^{*}_{31}, \ldots)$.

\subsection{M-model}

\citet{BotellaRocamoraetal:2015} describe a unifying modeling framework for
multivariate disease mapping when the number of diseases is potentially large.
Here, spatial effects are linear combinations of proper CAR spatial effects. In
particular, we will consider $K$ underlying proper CAR spatial effects defined by

$$ \phi_k \sim N\left(0, (D - \alpha_k W )^{-1}\right),\ k=1,\ldots,K.
$$ 
\noindent
Here, $\phi_k$ is a vector of length $I$.

The value of the spatial random effect $\Theta_{\cdot j}$ for variable $j$ is
defined as

$$
\Theta_{\cdot j} = \phi_1 m_{1j} + \ldots \phi_J m_{Jj} .
$$
\noindent
Hence, matrix $M$ with entries $m_{ij}$ defines the loadings of
the different underlying CAR spatial effects for each disease or variable.

The distribution of these random effects is given by

$$
vec(\Theta) \sim N\left(0, (M^{\top} \otimes I) 
  diag((\Sigma_w)_1, \ldots, (\Sigma_w)_K) (M \otimes I)\right) .
$$
\noindent
Here, matrices $(\Sigma_w)_k$ are the variance matrices of the $K$
underlying proper CAR spatial effects, i.e.,

$$
(\Sigma_w)_k = (D - \alpha_k W)^{-1},\ k=1,\ldots,K.
$$

In addition, \citet{BotellaRocamoraetal:2015} show that for the separable case, with $\alpha_1 = \ldots = \alpha_K$, the between-variables
variance matrix is $M^{\top} M$. The prior of this model is
on $M^{\top} M$, and it follows a Wishart with parameters $K$ and
$\tau I$. Parameter $\tau$ is a fixed precision which is set to
$0.001$, but it could also be considered as another hyperparameter to be
estimated \citep{CorpasBurgosetal:2019}.

The vector of hyperparameters in this model is made of the $K$ autocorrelation
parameters (conveniently transformed) followed by the columns of matrix
$M$, for which no transformation is required.

\subsection{Summary of the models}
\label{sec:summary}

To summarize the different models implemented and their associated functions,
Table~\ref{tab:models} displays some basic information about them.
Note how all functions names follow a similar pattern for consistency,
and that the latent models differ in the number of hyperparameters. These
depend on the number of variables $K$. Furthermore, there are two different
functions to define the new latent effects: the one that implements the
latent effects using the \code{rgeneric} framework and a wrapper function.
The wrapper function is intended to provide a simpler way to define
the latent effect and avoid calling function \code{inla.rgeneric.define}
\citep[see, for example, Chapter 11 in][]{INLAbook:2020}. This will be further
discussed in the examples in Section~\ref{sec:examples} below.

\begin{table} [h]
\centering
\begin{scriptsize}
\begin{tabular}{|c|c|c|c|}
\hline
Latent effect & Wrapper function & \code{rgeneric} function & \# hyperparameters \\
\hline
  Independent IMCAR  &  \code{inla.INDIMCAR.model} & \code{inla.rgeneric.indep.IMCAR.model}  &  $K$   \\
  Independent PMCAR  &  \code{inla.INDMCAR.model} & \code{inla.rgeneric.indep.MCAR.model}   &  $K+1$   \\
      IMCAR     &  \code{inla.IMCAR.model} &    \code{inla.rgeneric.IMCAR.model}      &  $K*(K+1)/2$   \\
      PMCAR     &  \code{inla.MCAR.model} &    \code{inla.rgeneric.MCAR.model}     &  $K*(K+1)/2 + 1$   \\
M-model         & \code{inla.Mmodel.model} & \code{inla.rgeneric.Mmodel.model}         &  $K * K$ \\
\hline
\end{tabular}
\end{scriptsize}
\caption{Summary of the latent effects implemented in the \pkg{INLAMSM}
 package.}
\label{tab:models}
\end{table}

\section{Examples}
\label{sec:examples}

In this section, two examples are developed with the \pkg{INLAMSM} package. The
first one is on the well-known North Carolina sudden infant death syndrome (SIDS) data
\citep{CressieRead:1985}, which is used to show how the different models
implemented in the \pkg{INLAMSM} package are fit. This example is also intended
to show that multivariate spatial models can not only be used to model several
diseases but also the same disease across different time periods so that
dependence across different periods can be estimated. The second example is
based on the study of three causes of death in Comunidad Valenciana (Spain), a
region that comprises 540 municipalities and provides a more challenging
dataset to fit spatial models than the North Carolina SIDS data (with only 100
areas).  In this second example, the focus in on investigating dependence among
the different diseases.

\subsection{North Carolina SIDS data}

In the first example, the North Carolina SIDS data \citep{CressieRead:1985}
have been considered to test the methods implemented in the \pkg{INLAMSM}
package.  This dataset includes counts of number of live births and number of
deaths from sudden infant death syndrome for the 100 counties of North
Carolina for two time periods: from July 1st, 1974 to Jun 30th, 1978 and from
July 1st, 1979 to June 30th, 1984.  This dataset is available in a shapefile in
\proglang{R} package \pkg{spData} \citep{spData}, which will be loaded using
package \pkg{rgdal} \citep{rgdal}.

\begin{Schunk}
\begin{Sinput}
R> library("rgdal")
R> #Load SIDS data
R> nc.sids <- readOGR(system.file("shapes/sids.shp", package = "spData")[1],
+    verbose = FALSE)
R> proj4string(nc.sids) <- CRS("+proj=longlat +ellps=clrk66")
\end{Sinput}
\end{Schunk}

Next, the adjacency matrix of the 100 counties in North Carolina
is obtained with function \code{poly2nb} in the \pkg{spdep} package
 \citep{AppliedVirgilio,spdep} and converted into a sparse matrix of type \code{Matrix} 
\citep{Matrix}.

\begin{Schunk}
\begin{Sinput}
R> library("spdep")
R> #Compute adjacency matrix, as nb object 'adj' and sparse matrix 'W'
R> adj <- poly2nb(nc.sids)
R> W <- as(nb2mat(adj, style = "B"), "Matrix")
\end{Sinput}
\end{Schunk}

The model that will be fit in this case is

$$
O_{ik} \sim Po(\mu_{ik});\
\log(\mu_{ik}) = \log(E_{ik}) + a_k + \theta_{ik},\ i=1,\ldots,100,\ k=1,2
$$
\noindent
Here, $O_{ik}$ is the number of SIDS cases in county $i$ and time period $k$,
$E_{ik}$ the expected counts, $a_k$ a period-specific intercept and
$\theta_{ik}$ the multivariate spatial effect, which is defined using one of
the models implemented in the \pkg{INLAMSM} package.
Note that other covariates and effects could be included in the linear
predictor as well.

The expected number of cases $E_{ik}$ are computed by multiplying the
period-specific mortality rate $r_k$ by the number of births $N_{ik}$:

$$
E_{ik} = r_k N_{ik};\ r_k = \frac{\sum_{i=1}^{100} O_{ik}}{\sum_{i=1}^{100} N_{ik}},\ i=1,\ldots,I,\ k=1,2 .
$$

In the next lines of \proglang{R} code
the expected counts for both time periods are computed, as well as the
standardized mortality ratio (SMR), $O_{ik} / E_{ik}$, and the proportion of
non-white births, which could be used as a covariate for both time periods.

\begin{Schunk}
\begin{Sinput}
R> # First time period
R> # Compute expected cases
R> r74 <- sum(nc.sids$SID74) / sum(nc.sids$BIR74)
R> nc.sids$EXP74 <- r74 * nc.sids$BIR74
R> # SMR
R> nc.sids$SMR74 <- nc.sids$SID74 / nc.sids$EXP74
R> # Proportion of non-white births
R> nc.sids$NWPROP74 <- nc.sids$NWBIR74 / nc.sids$BIR74
R> # Second time period
R> # Compute expected cases
R> r79 <- sum(nc.sids$SID79) / sum(nc.sids$BIR79)
R> nc.sids$EXP79 <- r79 * nc.sids$BIR79
R> # SMR
R> nc.sids$SMR79 <- nc.sids$SID79 / nc.sids$EXP79
R> # Proportion of non-white births
R> nc.sids$NWPROP79 <- nc.sids$NWBIR79 / nc.sids$BIR79
\end{Sinput}
\end{Schunk}

In order to prepare the data to fit the models with \pkg{INLA}, a new object is
created by appending the data from the first time period to the data from the
second one. In addition, to the observed, expected and proportion of non-white
births, and index is created to identify the counties and time periods.

\begin{Schunk}
\begin{Sinput}
R> d <- data.frame(OBS = c(nc.sids$SID74, nc.sids$SID79),
+    PERIOD = c(rep("74", 100), rep("79", 100)), 
+    NWPROP = c(nc.sids$NWPROP74, nc.sids$NWPROP79),
+    EXP = c(nc.sids$EXP74, nc.sids$EXP79))
R> # County-period index
R> d$idx <- 1:length(d$OBS)
\end{Sinput}
\end{Schunk}

Now, we will fit the models. This requires a two step process:

\begin{enumerate}

\item Define the multivariate spatial latent effect using one of the functions
in the \pkg{INLAMSM} package \citep{INLAMSM}. This is done via the
\code{inla.rgeneric.define} function, that will take the function of the
multivariate spatial model to be included plus any other required arguments,
such as the adjacency matrix, the number of time periods, etc.  Alternatively,
the associated wrapper function can be used (see below).

\item Fit the model with \pkg{INLA} \citep{INLApkg} using a formula that includes
the newly defined latent effect.

\end{enumerate}

For example, for the independent IMCAR model the latent effect will be 
defined as follows:

\begin{Schunk}
\begin{Sinput}
R> library("INLAMSM")
R> library("INLA")
R> # Number of variables (i.e., periods)
R> k <- 2
R> # Define bivariate latent effect
R> model.indimcar <- inla.rgeneric.define(inla.rgeneric.indep.IMCAR.model,
+    list(k = k, W = W))
\end{Sinput}
\end{Schunk}

The previous definition of the latent effect uses a call to function
\code{inla.rgeneric.define}, which takes function
\code{inla.rgeneric.indep.IMCAR.model} that defines the new latent effect plus
any other required arguments (i.e., a named list with the number of diseases
\code{k} and adjacency matrix \code{W}).  However, as explained in
Section~\ref{sec:summary} there are wrapper functions associated to each new
latent effect, that will be used from now on. For this particular case, the
definition of the new latent effect using a wrapper function is:

\begin{Schunk}
\begin{Sinput}
R> model.indimcar <- inla.INDIMCAR.model(k = k, W = W)
\end{Sinput}
\end{Schunk}

As explained in Section~\ref{sec:models}, given that this model includes
intrinsic CAR specifications, it is necessary to include some constraints so
that the random effects for each disease sum up to zero. Constraints in
\pkg{INLA} are added using equation $\mathbf{A} \mathbf{x}^{\top} =
\mathbf{e}^{\top}$, where matrix $\mathbf{A}$ and vector $\mathbf{e}$ are used
to define linear constraints on the (multivariate) vector of random effects
$\mathbf{x}$. These two elements are defined below:

\begin{Schunk}
\begin{Sinput}
R> A <- kronecker(Diagonal(k, 1), Matrix(1, ncol = nrow(W), nrow = 1))
R> e  = rep(0, k)
\end{Sinput}
\end{Schunk}

\noindent
The constraints are specified using argument \code{extraconstr} in the call to
the \code{f} function that defines the latent effect (see below).

The next step fits the model with \pkg{INLA}:

\begin{Schunk}
\begin{Sinput}
R> #Fit model
R> IIMCAR <- inla(OBS ~ 0 + PERIOD + f(idx, model = model.indimcar,
+      extraconstr = list(A = as.matrix(A), e = e)),
+    data = d,
+    E = EXP, family = "poisson", control.predictor = list(compute = TRUE),
+    control.compute = list(dic = TRUE, waic = TRUE))
\end{Sinput}
\end{Schunk}

Note that linear constraints have been added so that random effects associated
to each disease sum up to zero, as explained above.  Furthermore, argument
\code{control.predictor} has been set to compute the marginals of the linear
predictor (i.e., by setting \code{compute} equal to \code{TRUE}) and argument
\code{control.compute} has been set to compute two model selection criteria. In
particular, the deviance information criterion
\citep[DIC]{spiegelhalteretal:2002} and Watanabe-Akaike information criterion
\citep[WAIC]{Watanabe:2013} will be computed. See below for a discussion on how
to use these criteria for model choice.

This model can be summarized as follows: 

\begin{Schunk}
\begin{Sinput}
R> summary(IIMCAR)
\end{Sinput}
\begin{Soutput}
Call:
   c("inla(formula = OBS ~ 0 + PERIOD + f(idx, model = 
   model.indimcar, ", " extraconstr = list(A = as.matrix(A), e = 
   e)), family = \"poisson\", ", " data = d, E = EXP, 
   control.compute = list(dic = TRUE, waic = TRUE), ", " 
   control.predictor = list(compute = TRUE))") 
Time used:
    Pre = 1.21, Running = 3.51, Post = 0.1, Total = 4.82 
Fixed effects:
           mean    sd 0.025quant 0.5quant 0.975quant   mode kld
PERIOD74 -0.071 0.056     -0.184   -0.070      0.036 -0.067   0
PERIOD79 -0.017 0.046     -0.110   -0.016      0.071 -0.015   0

Random effects:
  Name	  Model
    idx RGeneric2

Model hyperparameters:
                mean    sd 0.025quant 0.5quant 0.975quant  mode
Theta1 for idx 0.774 0.339      0.129    0.764       1.46 0.731
Theta2 for idx 1.429 0.408      0.664    1.414       2.27 1.361

Expected number of effective parameters(stdev): 65.92(8.94)
Number of equivalent replicates : 3.03 

Deviance Information Criterion (DIC) ...............: 909.59
Deviance Information Criterion (DIC, saturated) ....: 271.91
Effective number of parameters .....................: 67.35

Watanabe-Akaike information criterion (WAIC) ...: 910.58
Effective number of parameters .................: 53.61

Marginal log-Likelihood:  -484.51 
Posterior marginals for the linear predictor and
 the fitted values are computed
\end{Soutput}
\end{Schunk}

Note that hyperparameters \code{Theta1} and \code{Theta2} correspond to the
log-precisions of the latent effect, i.e., the two independent ICAR spatial
effects. Because the two effects are independent of each other, this
model is equivalent to fitting two models using an intrinsic CAR model
(implemented as the \code{besag} latent effect in \pkg{INLA}).

\begin{figure}
\centering
\includegraphics{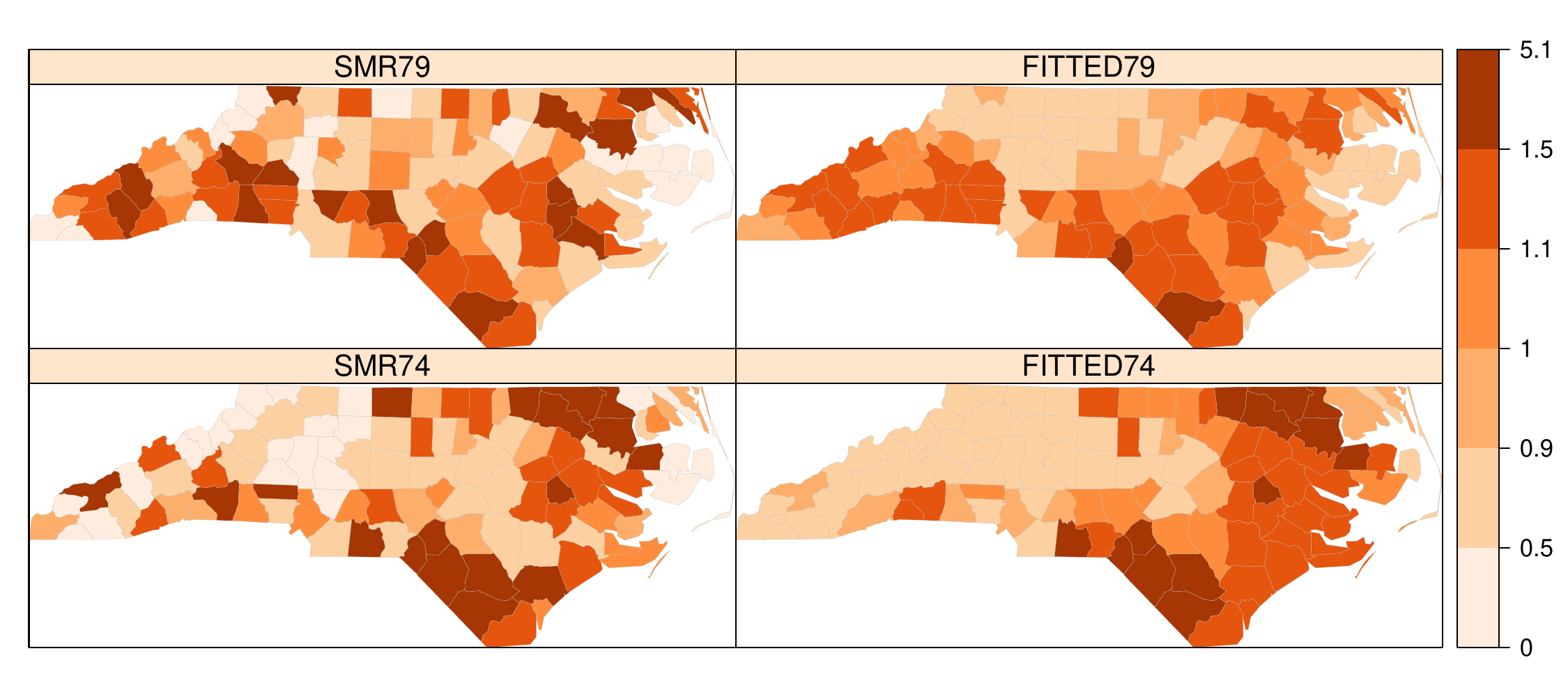}
\caption{Standardized mortality ratio and posterior means of the fitted values with the IIMCAR model for both time periods for the North Carolina SIDS data.}
\label{fig:ncsids}
\end{figure}

Next, the fitted values of the relative risks of the IIMCAR model can be
compared with the SMR for both time periods as seen in Figure~\ref{fig:ncsids}.
Package \pkg{RColorBrewer} \citep{RColorBrewer} has been used to set the colors
in the palette. Note how the fitted values show smoothed spatial patterns as
compared with the raw SMR values.

The other models implemented in the \pkg{INLAMSM} package can be fit in a
similar way, as listed below. For example, the independent proper MCAR model
requires the values of $\alpha_{min}$ and $\alpha_{max}$ to be passed in
addition to \code{k} and \code{W} when defining the latent effect. This is
exemplified below. We have constrained $\alpha$ to take just positive values
since the PCAR distribution has been said to have a counterintuitive
performance when this parameter takes negative values \citep{Wall:2004}.
Anyway, our implementation of the PCAR distribution admits, if wanted, the
whole range of admissible values for this parameter \citep{Sunetal:1999}.

\begin{Schunk}
\begin{Sinput}
R> # Independent Proper MCAR model
R> # Define range for the autocorrelation parameter
R> alpha.min <- 0
R> alpha.max <- 1
R> model.indmcar <- inla.INDMCAR.model(k = k, W = W, alpha.min = alpha.min,
+    alpha.max = alpha.max)
R> #Fit model
R> IPMCAR <- inla(OBS ~ 0 + PERIOD + f(idx, model = model.indmcar), data = d,
+    E = EXP, family = "poisson", control.predictor = list(compute = TRUE),
+    control.compute = list(dic = TRUE, waic = TRUE))
\end{Sinput}
\end{Schunk}

The remainder of the models in the \pkg{INLAMSM} package are defined and
fit as:

\begin{Schunk}
\begin{Sinput}
R> # IMCAR model
R> model.imcar <- inla.IMCAR.model(k = k, W = W)
R> #Fit model
R> IMCAR <- inla(OBS ~ 0 + PERIOD + f(idx, model = model.imcar,
+      extraconstr = list(A = as.matrix(A), e = e)),
+    data = d,
+    E = EXP, family = "poisson", control.predictor = list(compute = TRUE),
+    control.compute = list(dic = TRUE, waic = TRUE))
\end{Sinput}
\end{Schunk}

As discussed above, linear constraints have been added when fitting this model
too.  This way, the intercepts in the model can be identified as well.

\begin{Schunk}
\begin{Sinput}
R> # Proper MCAR model
R> model.mcar <- inla.MCAR.model(k = k, W = W, alpha.min = alpha.min,
+    alpha.max = alpha.max)
R> #Fit model
R> PMCAR <- inla(OBS ~ 0 + PERIOD + f(idx, model = model.mcar), data = d,
+    E = EXP, family = "poisson", control.predictor = list(compute = TRUE),
+    control.compute = list(dic = TRUE, waic = TRUE))
\end{Sinput}
\end{Schunk}

\begin{Schunk}
\begin{Sinput}
R> # M-model
R> model.mmodel <- inla.Mmodel.model(k = k, W = W, alpha.min = alpha.min,
+    alpha.max = alpha.max)
R> # Fit model
R> Mmodel <- inla(OBS ~ 0 + PERIOD + f(idx, model = model.mmodel), data = d,
+    E = EXP, family = "poisson", control.predictor = list(compute = TRUE),
+    control.compute = list(dic = TRUE, waic = TRUE))
\end{Sinput}
\end{Schunk}

\begin{figure}
\centering
\includegraphics{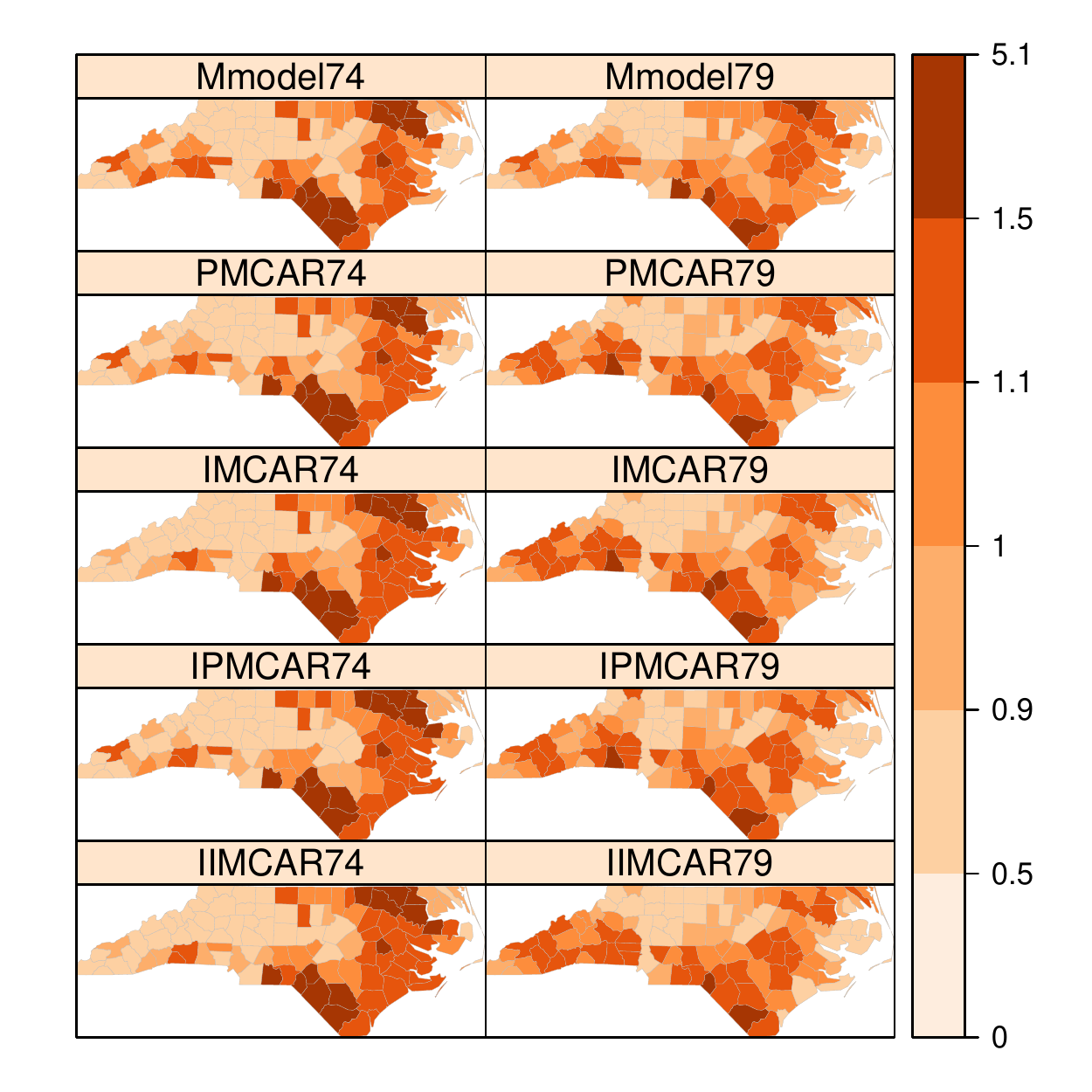}
\caption{Posterior means of the relative risks for the different multivariate spatial effects fit to the North Carolina SIDS data.}
\label{fig:ncsidsmsm}
\end{figure}

Once we have fit all models, we can compare point estimates of the relative
risks. Posterior means of the relative risks from different models are
displayed in Figure~\ref{fig:ncsidsmsm}.

\begin{Schunk}
\begin{Sinput}
R> # Add results to nc.sids
R> nc.sids$IIMCAR74 <- IIMCAR$summary.fitted[1:100, "mean"]
R> nc.sids$IIMCAR79 <- IIMCAR$summary.fitted[100 + 1:100, "mean"]
R> nc.sids$IPMCAR74 <- IPMCAR$summary.fitted[1:100, "mean"]
R> nc.sids$IPMCAR79 <- IPMCAR$summary.fitted[100 + 1:100, "mean"]
R> nc.sids$IMCAR74 <- IMCAR$summary.fitted[1:100, "mean"]
R> nc.sids$IMCAR79 <- IMCAR$summary.fitted[100 + 1:100, "mean"]
R> nc.sids$PMCAR74 <- PMCAR$summary.fitted[1:100, "mean"]
R> nc.sids$PMCAR79 <- PMCAR$summary.fitted[100 + 1:100, "mean"]
R> nc.sids$Mmodel74 <- Mmodel$summary.fitted[1:100, "mean"]
R> nc.sids$Mmodel79 <- Mmodel$summary.fitted[100 + 1:100, "mean"]
\end{Sinput}
\end{Schunk}

\subsubsection{Including covariates in the model}

\begin{figure}
\centering
\includegraphics{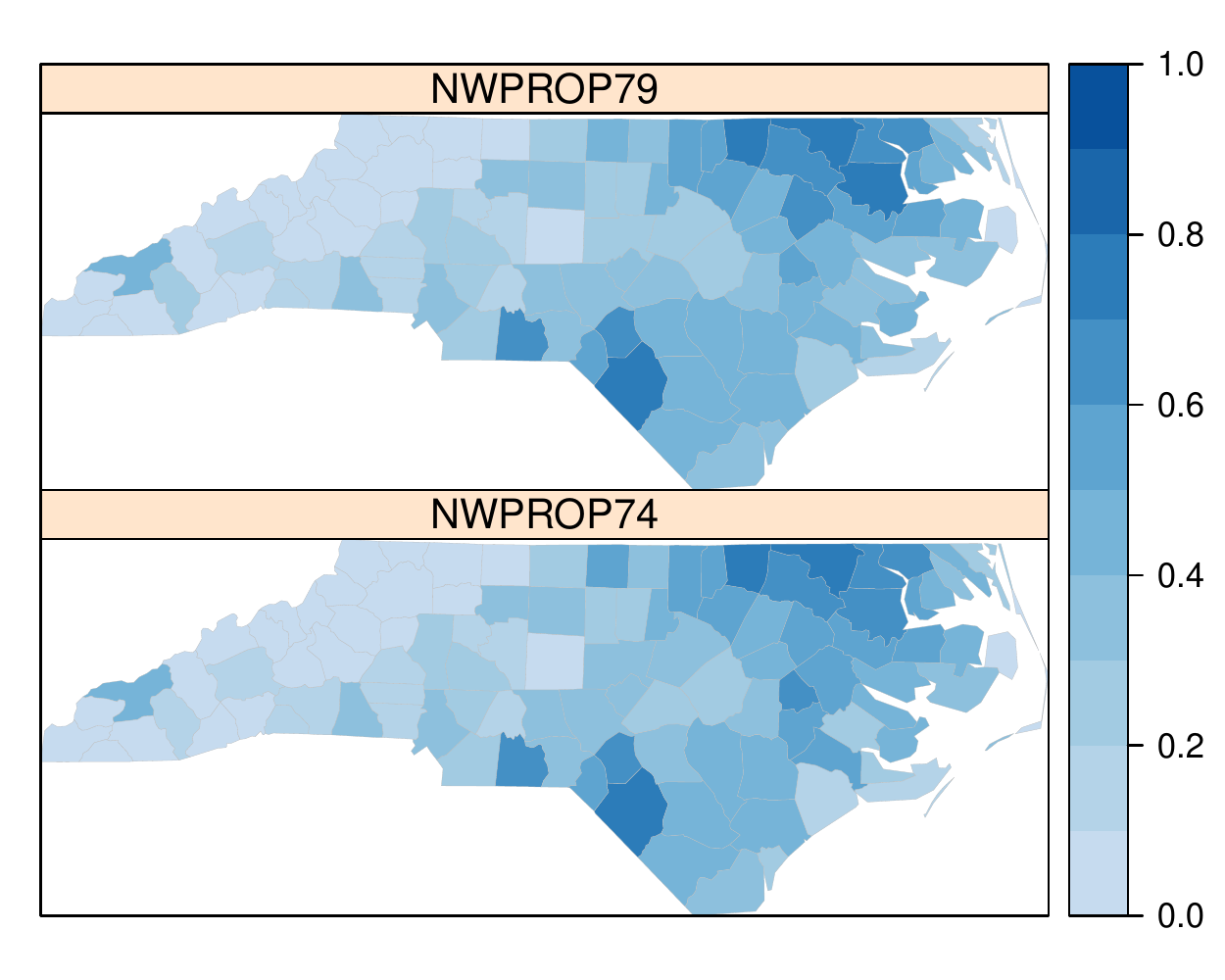}
\caption{Proportion of non-white births per county for both time periods.}
\label{fig:nwprop}
\end{figure}

The proportion of non-white births shows a very similar pattern to that of the
relative risk. These proportions for the time periods are shown in
Figure~\ref{fig:nwprop}, which can be compared with the estimates of the
relative risk in Figure~\ref{fig:ncsids} to appreciate the similar patterns.
Note the similar patterns between the proportion of non-white births and the
standardized mortality ratio for both time periods.  For this reason, several
authors \citep[see, for example,][]{CressieRead:1985} have mentioned the
importance of including this covariate in the model. Hence, the same models are
fitted now including the covariate so that a different coefficient is estimated
for each time period.

In order to have the covariate in the model so that a different coefficient is
estimated  for each time period, the covariate must be structured in a
two-column as follows:

\begin{Schunk}
\begin{Sinput}
R> # Number of areas
R> n <- nrow(W)
R> NWPROP <- matrix(NA, ncol = 2, nrow = 2 * n)
R> NWPROP[1:n, 1] <- nc.sids$NWPROP74
R> NWPROP[n + 1:n, 2] <- nc.sids$NWPROP79
\end{Sinput}
\end{Schunk}

Next, data used to fit the previous models are converted into a list so that
the covariate can be added:

\begin{Schunk}
\begin{Sinput}
R> d <- as.list(d)
R> d$NWPROP <- NWPROP
\end{Sinput}
\end{Schunk}

Finally, models are fit again:

\begin{Schunk}
\begin{Sinput}
R> # Independent ICAR model
R> IIMCAR2 <- inla(OBS ~ 0 + PERIOD + NWPROP + f(idx, model = model.indimcar,
+      extraconstr = list(A = as.matrix(A), e = e)),
+    data = d,
+    E = EXP, family = "poisson", control.predictor = list(compute = TRUE),
+    control.compute = list(dic = TRUE, waic = TRUE))
\end{Sinput}
\end{Schunk}
\begin{Schunk}
\begin{Sinput}
R> # Independent proper CAR model
R> IPMCAR2 <- inla(OBS ~ 0 + PERIOD + NWPROP + f(idx, model = model.indmcar),
+    data = d,
+    E = EXP, family = "poisson", control.predictor = list(compute = TRUE),
+    control.compute = list(dic = TRUE, waic = TRUE))
\end{Sinput}
\end{Schunk}
\begin{Schunk}
\begin{Sinput}
R> # Instrinsic MCAR model
R> IMCAR2 <- inla(OBS ~ 0 + PERIOD + NWPROP + f(idx, model = model.imcar,
+      extraconstr = list(A = as.matrix(A), e = e)),
+    data = d,
+    E = EXP, family = "poisson", control.predictor = list(compute = TRUE),
+    control.compute = list(dic = TRUE, waic = TRUE))
\end{Sinput}
\end{Schunk}
\begin{Schunk}
\begin{Sinput}
R> # Proper MCAR model
R> PMCAR2 <- inla(OBS ~ 0 + PERIOD + NWPROP + f(idx, model = model.mcar),
+    data = d, E = EXP, family = "poisson", 
+    control.predictor = list(compute = TRUE),
+    control.compute = list(dic = TRUE, waic = TRUE))
\end{Sinput}
\end{Schunk}
\begin{Schunk}
\begin{Sinput}
R> # M-model
R> Mmodel2 <- inla(OBS ~ 0 + PERIOD + NWPROP + f(idx, model = model.mmodel),
+    data = d, E = EXP, family = "poisson",
+    control.predictor = list(compute = TRUE),
+    control.compute = list(dic = TRUE, waic = TRUE))
\end{Sinput}
\end{Schunk}

Note that, in a similar manner that sum-to-zero constraints have been imposed
for intrinsic CAR distributions, further constraints could be also imposed 
above assuming the spatial random effects to be orthogonal to the
covariate. In this manner, confounding between the covariate and spatial random
effects could be alleviated \citep{HodgesReich:2010}.

In order to show the estimates of the coefficients of the covariates, the
summary of the independent IMCAR model with covariates is shown below:

\begin{Schunk}
\begin{Sinput}
R> summary(IIMCAR2)
\end{Sinput}
\begin{Soutput}
Call:
   c("inla(formula = OBS ~ 0 + PERIOD + NWPROP + f(idx, model = 
   model.indimcar, ", " extraconstr = list(A = as.matrix(A), e = 
   e)), family = \"poisson\", ", " data = d, E = EXP, 
   control.compute = list(dic = TRUE, waic = TRUE), ", " 
   control.predictor = list(compute = TRUE))") 
Time used:
    Pre = 1.37, Running = 4.22, Post = 0.0786, Total = 5.67 
Fixed effects:
           mean    sd 0.025quant 0.5quant 0.975quant   mode kld
PERIOD74 -0.687 0.130     -0.949   -0.684     -0.438 -0.679   0
PERIOD79 -0.211 0.124     -0.458   -0.210      0.031 -0.208   0
NWPROP1   1.978 0.351      1.297    1.974      2.681  1.966   0
NWPROP2   0.622 0.365     -0.103    0.623      1.336  0.627   0

Random effects:
  Name	  Model
    idx RGeneric2

Model hyperparameters:
               mean    sd 0.025quant 0.5quant 0.975quant mode
Theta1 for idx 1.91 0.587      0.867     1.87       3.16 1.72
Theta2 for idx 1.54 0.437      0.733     1.53       2.44 1.46

Expected number of effective parameters(stdev): 49.28(9.58)
Number of equivalent replicates : 4.06 

Deviance Information Criterion (DIC) ...............: 899.49
Deviance Information Criterion (DIC, saturated) ....: 261.81
Effective number of parameters .....................: 51.51

Watanabe-Akaike information criterion (WAIC) ...: 904.67
Effective number of parameters .................: 46.14

Marginal log-Likelihood:  -479.10 
Posterior marginals for the linear predictor and
 the fitted values are computed
\end{Soutput}
\end{Schunk}

\begin{figure}
\centering
\includegraphics{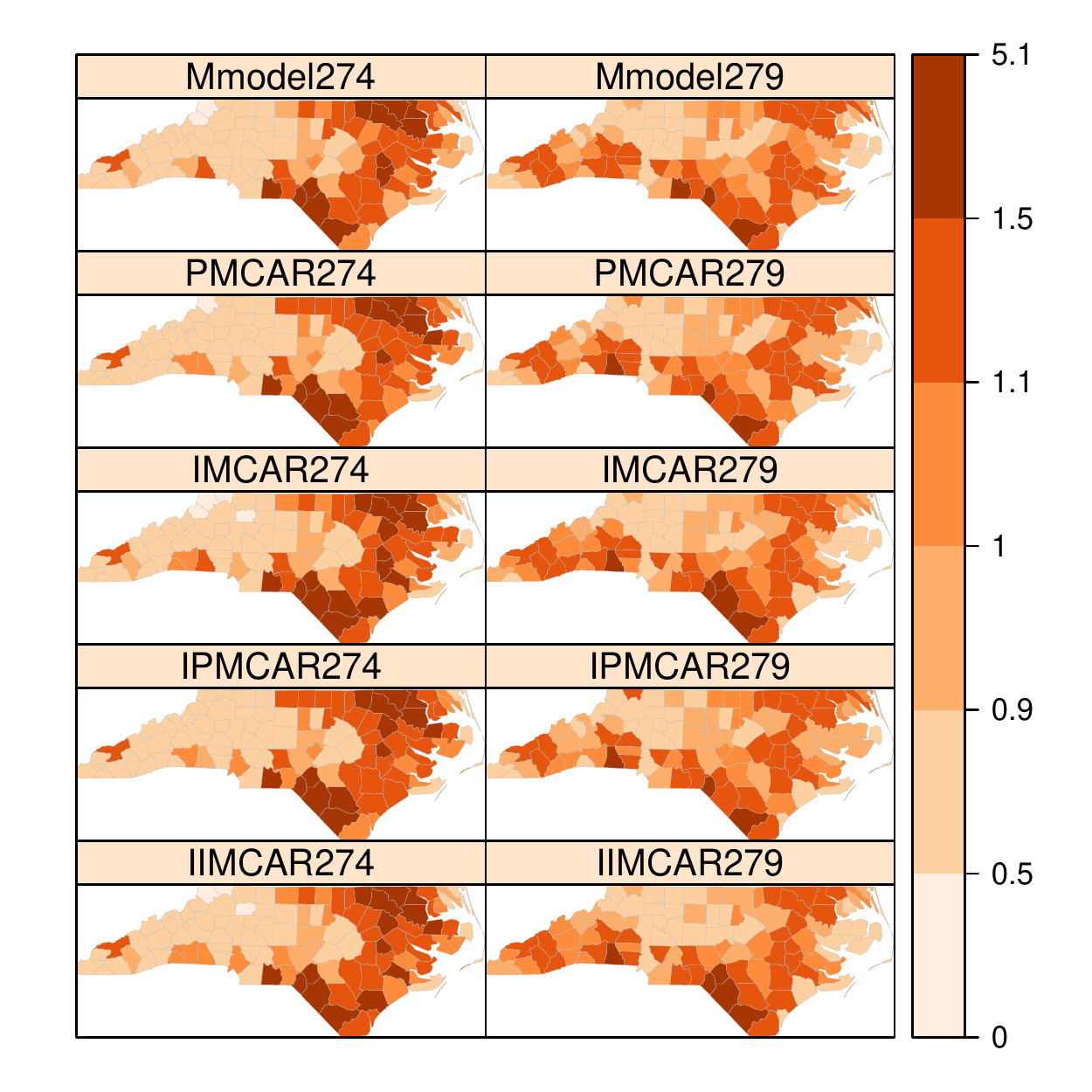}
\caption{Posterior means of the relative risks from the different multivariate
spatial effects fit to the North Carolina SIDS data. These models include the
proportion of non-white births as a covariate.}
\label{fig:ncsidsmsm2}
\end{figure}

These results show the positive association between the relative risk and the
proportion of non-white births. Other models have similar estimates of the
coefficients of this  covariate. In addition, Figure~\ref{fig:ncsidsmsm2} shows
the posterior means of the relative risks estimated with these models.  They
can be compared to the ones in Figure~\ref{fig:ncsidsmsm} to notice that the
estimates are usually very similar.

\subsubsection{Model selection}

Given the large number of models fit to the data it is necessary to select the
best model among all of them. The deviance information criterion
\citep[DIC]{spiegelhalteretal:2002} and Watanabe-Akaike information criterion
\citep[WAIC]{Watanabe:2013} have been computed when the models were fit and
they are shown in Table~\ref{tab:NCSIDSIC}.

\begin{table}[h!]
\centering
\begin{tabular}{c|c|c|c|c}
Model & \multicolumn{2}{c}{DIC} & \multicolumn{2}{|c}{WAIC} \\
\hline
 & \multicolumn{2}{c}{Covariate} & \multicolumn{2}{|c}{Covariate} \\
\cline{2-5}
 & No & Yes & No & Yes\\
\cline{2-5}
\hline
Independent IMCAR & 
909.59 & 899.49 & 
910.58 & 904.67 \\
Independent PMCAR & 
910.59 & 897.87 & 
903.75 & 895.61 \\
IMCAR &
904.26 & 894.96 & 
905.96 & 897.38 \\
PMCAR & 
905.47 & 893.8 & 
898.34 & 888.76 \\
M-model &
911.47 & 893.3 & 
917.71 & 898.17 \\
\end{tabular}
\caption{Information criteria computed for model selection.}
\label{tab:NCSIDSIC}
\end{table}

According to the values obtained of the DIC and WAIC, it is clear that
the covariate should be included in the model because it produces a decrease
in the DIC and WAIC of about 10 for all models. Secondly, models with 
independent effects
in general do not perform as well as the other models. This means that
there is some within-area association between the different variables that
needs to be accounted for. 

\subsection{Mortality in Comunidad Valenciana (Spain)}

The next example is based on simulated data of the mortality by cirrhosis, lung
and oral cancer in Comunidad Valenciana (Spain).  This dataset has been
obtained from \citet{bookMSM:2019} and it has been generated to mimic the
spatial pattern of the real data, that cannot be provided due to
confidentiality constraints. The original files are available at
\texttt{http://github.com/MigueBeneito/DisMapBook}.

Here, the number of deaths by these three causes are available at the
municipality level in Comunidad Valencia (Spain), as well as the expected
number of cases that have been computed using internal standardization.
Hence, the aim now is to estimate the spatial pattern of the different
diseases as well as their possible correlations.

Given that in the previous example we have already described how to fit
all the new models in the \proglang{R} package, we will focus now on the models
that include a term to model the covariance for several diseases. That is,
only models IMCAR, PMCAR and M-model will be fit.

Comunidad Valenciana data are available in the package as two separate objects:
\code{CV} and \code{CV.nb}. \code{CV} is a \code{SpatialPolygonsDataFrame}
\citep{AppliedVirgilio} that contains the data and boundaries of the
municipalities in Comunidad Valenciana, while \code{CV.nb} is an \code{nb}
object \citep{spdep} with the neighborhood structure. Both objects can be
loaded as:

\begin{Schunk}
\begin{Sinput}
R> #Load data
R> data(CV)
\end{Sinput}
\end{Schunk}

The adjacency matrix is computed using a sparse representation \citep{Matrix}
as follows:

\begin{Schunk}
\begin{Sinput}
R> #Compute sparse adjacency matrix W
R> W <- as(nb2mat(CV.nb, style = "B"), "Matrix")
\end{Sinput}
\end{Schunk}

Next, a \code{data.frame} is created with the observed and expected data,
together with an index variable to be passed to the definition of the latent
effects and a disease-specific intercept: 

\begin{Schunk}
\begin{Sinput}
R> #Data
R> d <- data.frame(OBS = c(CV$Obs.Cirrhosis, CV$Obs.Lung, CV$Obs.Oral),
+    EXP = c(CV$Exp.Cirrhosis, CV$Exp.Lung, CV$Exp.Oral)
+  )
R> # Add disease specific intercept
R> d$Intercept <- rep(c("Cirrhosis", "Lung", "Oral"), each = nrow(W))
R> # Index for latent effect
R> d$idx <- 1:length(d$OBS)
\end{Sinput}
\end{Schunk}

In order to fit the different models, the following parameters
are defined:

\begin{Schunk}
\begin{Sinput}
R> #Number of diseases
R> k <- 3 
R> # Range of autocorrelation parameter
R> alpha.min <- 0
R> alpha.max <- 1
\end{Sinput}
\end{Schunk}

Similarly to the SIDS example, some of the multivariate spatial
effects require additional constraints. The matrices required
are defined below:

\begin{Schunk}
\begin{Sinput}
R> A <- kronecker(Diagonal(k, 1), Matrix(1, ncol = nrow(W), nrow = 1))
R> e = rep(0, k)
\end{Sinput}
\end{Schunk}

Then, the latent effects are defined and the models are fit:

\begin{Schunk}
\begin{Sinput}
R> # Define latent IMCAR model
R> model <- inla.IMCAR.model(k = k, W = W)
R> # FIT IMCAR model
R> IMCAR.cval <- inla(OBS ~ 0 + Intercept + f(idx, model = model,
+      extraconstr = list(A = as.matrix(A), e = e)),
+    data = d, E = EXP, family = "poisson",
+    control.compute = list(config = TRUE, dic = TRUE, waic = TRUE),
+    # Increase starting values of marginal precisions for model fitting
+    control.mode = list(theta = c(1, 1, 1, 0, 0, 0), restart = TRUE),
+    # Required to obtain more accurate gradients for model fitting
+    control.inla = list(h = 0.001),
+    control.predictor = list(compute = TRUE))
R> # Define latent PMCAR model
R> model <- inla.MCAR.model(k = k, W = W, alpha.min = alpha.min,
+    alpha.max = alpha.max)
R> # Fit PMCAR model
R> PMCAR.cval <-  inla(OBS ~ 0 + Intercept + f(idx, model = model), data = d,
+    E = EXP, family = "poisson",
+    control.compute = list(config = TRUE, dic = TRUE, waic = TRUE),
+    control.predictor = list(compute = TRUE))
R> # Define latent M-model
R> model <- inla.Mmodel.model(k = k, W = W, alpha.min = alpha.min,
+    alpha.max = alpha.max)
R> # Fit M-model
R> Mmodel.cval <- inla(OBS ~ 0 + Intercept + f(idx, model = model), data = d,
+    E = EXP, family = "poisson",
+    control.compute = list(config = TRUE, dic = TRUE, waic = TRUE),
+    control.predictor = list(compute = TRUE)
+  )
\end{Sinput}
\end{Schunk}

Table~\ref{tab:times} shows the computing times for the models fit to both
examples. All models have been run on a Mac OS X computer with an Intel Core i5
processor (2,7 GHz), 4 cores and 16GB of RAM. Now the models take longer to run
than in the previous example because there are three diseases and about 5 times
more areas (as Comunidad Valenciana has 540 municipalities in the cartography
that we have used). \citet{BotellaRocamoraetal:2015} report times of about 16
minutes to fit the M-model with \pkg{WinBUGS} on this same dataset.
Hence, \pkg{INLA} can fit the same models in a fraction of the
time.

In addition, Table~\ref{tab:times} also shows the computing times of
models fit to the Comunidad Valenciana dataset using two diseases (cirrhosis
and lung cancer). These models have been computed as follows:

\begin{Schunk}
\begin{Sinput}
R> # New dataset
R> d2 <- subset(d,  Intercept != "Oral")
R> d2$idx <- 1:nrow(d2)
R> k <- 2
R> # Linear constraint for the new models
R> A <- kronecker(Diagonal(k, 1), Matrix(1, ncol = nrow(W), nrow = 1))
R> e = rep(0, k)
R> model <- inla.IMCAR.model(k = k, W = W)
R> # FIT IMCAR model
R> IMCAR.cval2 <- inla(OBS ~ 0 + Intercept + f(idx, model = model,
+      extraconstr = list(A = as.matrix(A), e = e)),
+    data = d2, E = EXP, family = "poisson",
+    control.compute = list(config = TRUE, dic = TRUE, waic = TRUE),
+    control.predictor = list(compute = TRUE))
R> # Define latent PMCAR model
R> model <- inla.MCAR.model(k = k, W = W, alpha.min = alpha.min,
+    alpha.max = alpha.max)
R> # Fit PMCAR model
R> PMCAR.cval2 <-  inla(OBS ~ 0 + Intercept + f(idx, model = model),
+    data = d2, E = EXP, family = "poisson",
+    control.compute = list(config = TRUE, dic = TRUE, waic = TRUE),
+    control.predictor = list(compute = TRUE))
R> # Define latent M-model
R> model <- inla.Mmodel.model(k = k, W = W, alpha.min = alpha.min,
+    alpha.max = alpha.max)
R> # Fit M-model
R> Mmodel.cval2 <- inla(OBS ~ 0 + Intercept + f(idx, model = model),
+    data = d2, E = EXP, family = "poisson",
+    control.compute = list(config = TRUE, dic = TRUE, waic = TRUE),
+    control.predictor = list(compute = TRUE)
+  )
\end{Sinput}
\end{Schunk}

\begin{table}[h]
\begin{scriptsize}
\begin{tabular}{|c|c|c|c|c|c|c|c|}
\hline
 &  &  & \multicolumn{5}{c|}{Model} \\
\cline{4-8}
Dataset & \# diseases & \# Areas & Ind. IMCAR & Ind. PMCAR & IMCAR & PMCAR & M-model\\
\hline
NC SIDS &  2 & 100 & 
  4.82 & 
  3.72 & 
  8.88 & 
  9.19 & 
  15.64\\
C. Valenciana &  2 & 540 & -- & -- & 
  10.5 & 
  16.92 & 
  62.41\\
C. Valenciana &  3 & 540 & -- & -- & 
  36.51 & 
  45.67 & 
  319.46\\
\hline
\end{tabular}
\end{scriptsize}
\caption{Computing times (in seconds) for the models fit to the two examples.}
\label{tab:times}
\end{table}

The fitted values (i.e., posterior means of the relative risks) can be added to
the map in the \code{SpatialPolygonsDataFrame} object as follows:

\begin{Schunk}
\begin{Sinput}
R> n <- nrow(W)
R> CV$IMCAR.CIR <- IMCAR.cval$summary.fitted[1:n, "mean"] 
R> CV$IMCAR.LUN <- IMCAR.cval$summary.fitted[n + 1:n, "mean"] 
R> CV$IMCAR.ORA <- IMCAR.cval$summary.fitted[2 * n + 1:n, "mean"] 
R> CV$PMCAR.CIR <- PMCAR.cval$summary.fitted[1:n, "mean"] 
R> CV$PMCAR.LUN <- PMCAR.cval$summary.fitted[n + 1:n, "mean"] 
R> CV$PMCAR.ORA <- PMCAR.cval$summary.fitted[2 * n + 1:n, "mean"] 
R> CV$Mmodel.CIR <- Mmodel.cval$summary.fitted[1:n, "mean"] 
R> CV$Mmodel.LUN <- Mmodel.cval$summary.fitted[n + 1:n, "mean"] 
R> CV$Mmodel.ORA <- Mmodel.cval$summary.fitted[2 * n + 1:n, "mean"] 
\end{Sinput}
\end{Schunk}

\begin{figure}
\centering
\includegraphics{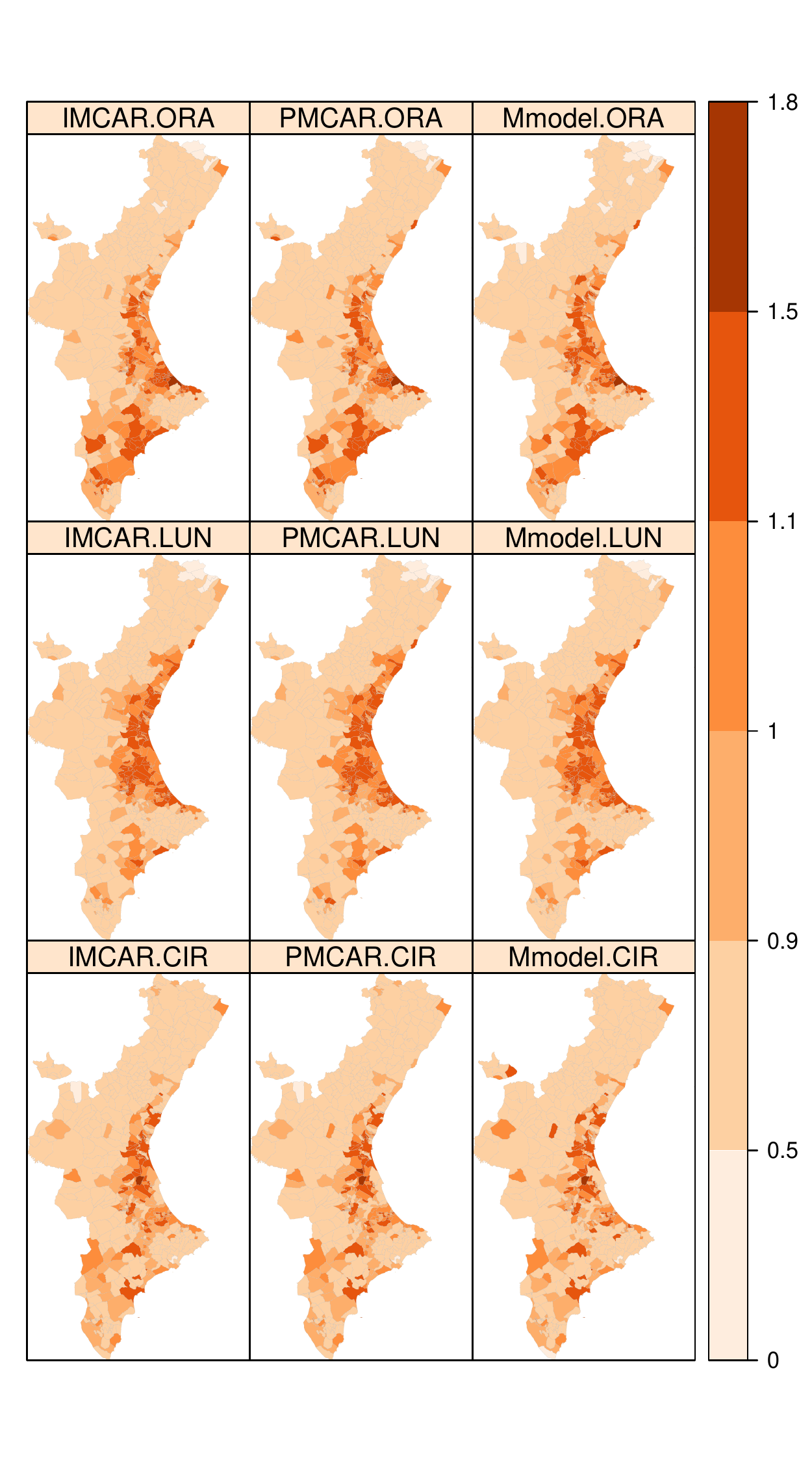}
\caption{Posterior means of the relative risks of cirrhosis, lung cancer and oral cancer in Comunidad Valenciana (Spain).}
\label{fig:maps}
\end{figure}

The maps in Figure~\ref{fig:maps} show the different posterior means of the
relative risks from the models fitted for the different causes of death. In
general, the three models produce similar point estimates of the relative
risks.

The \pkg{INLAMSM} package includes a few functions to transform the marginals
and summary statistics of the  model hyperparameters in the internal scale into
the original scale in the model. For the models fitted, this
back-transformation  can be obtained as follows:

\begin{Schunk}
\begin{Sinput}
R> hyper.imcar <- inla.MCAR.transform(IMCAR.cval, 3)
R> hyper.pmcar <- inla.MCAR.transform(PMCAR.cval, 3, model = "PMCAR",
+    alpha.min = alpha.min, alpha.max = alpha.max)
R> hyper.mmodel <- inla.Mmodel.transform(Mmodel.cval, 3,
+    alpha.min = alpha.min, alpha.max = alpha.max)
\end{Sinput}
\end{Schunk}

These will provide a transformation of the model parameters so that they are
not in the internal scale anymore and make inference easier.  Spatial
autocorrelation parameters are transformed to be in the range between
$\alpha_{min}$ and $\alpha_{max}$, and log-precisions are transformed to be
variances. Correlation hyperparameters are transformed to be between $-1$ and
$1$.

Hence, summary statistics for the models are:

\begin{Schunk}
\begin{Sinput}
R> # IMCAR
R> hyper.imcar$summary.hyperpar
\end{Sinput}
\begin{Soutput}
                mean     sd quant0.025 quant0.25 quant0.5 quant0.75
Theta1 for idx 0.243 0.0396     0.1752     0.214    0.239     0.267
Theta2 for idx 0.127 0.0162     0.0971     0.115    0.126     0.137
Theta3 for idx 0.226 0.0461     0.1490     0.193    0.222     0.254
Theta4 for idx 0.469 0.0829     0.2976     0.415    0.472     0.527
Theta5 for idx 0.563 0.0964     0.3542     0.502    0.571     0.632
Theta6 for idx 0.705 0.0633     0.5657     0.666    0.711     0.751
               quant0.975
Theta1 for idx      0.330
Theta2 for idx      0.160
Theta3 for idx      0.329
Theta4 for idx      0.622
Theta5 for idx      0.730
Theta6 for idx      0.813
\end{Soutput}
\begin{Sinput}
R> # PMCAR
R> hyper.pmcar$summary.hyperpar
\end{Sinput}
\begin{Soutput}
                mean      sd quant0.025 quant0.25 quant0.5 quant0.75
Theta1 for idx 0.989 0.00634      0.973     0.986    0.991     0.994
Theta2 for idx 0.262 0.04355      0.190     0.231    0.257     0.288
Theta3 for idx 0.139 0.01804      0.106     0.126    0.138     0.150
Theta4 for idx 0.254 0.05207      0.169     0.217    0.248     0.285
Theta5 for idx 0.458 0.08174      0.291     0.403    0.460     0.515
Theta6 for idx 0.545 0.09598      0.340     0.483    0.551     0.613
Theta7 for idx 0.688 0.06531      0.547     0.646    0.692     0.734
               quant0.975
Theta1 for idx      0.998
Theta2 for idx      0.361
Theta3 for idx      0.176
Theta4 for idx      0.372
Theta5 for idx      0.610
Theta6 for idx      0.715
Theta7 for idx      0.802
\end{Soutput}
\begin{Sinput}
R> #M-model
R> hyper.mmodel$summary.hyperpar
\end{Sinput}
\begin{Soutput}
                  mean      sd quant0.025 quant0.25 quant0.5
Theta1 for idx  0.2659 0.18179     0.0250    0.1169   0.2296
Theta2 for idx  0.9963 0.00335     0.9874    0.9950   0.9972
Theta3 for idx  0.9900 0.01186     0.9573    0.9875   0.9938
Theta4 for idx  0.5367 0.05211     0.4358    0.5011   0.5359
Theta5 for idx  0.1836 0.03697     0.1084    0.1590   0.1846
Theta6 for idx  0.1566 0.04985     0.0610    0.1224   0.1554
Theta7 for idx  0.0373 0.04748    -0.0546    0.0049   0.0366
Theta8 for idx  0.2946 0.02608     0.2423    0.2771   0.2951
Theta9 for idx  0.1820 0.04260     0.1007    0.1527   0.1809
Theta10 for idx 0.1292 0.08541    -0.0398    0.0715   0.1295
Theta11 for idx 0.2395 0.04345     0.1544    0.2100   0.2393
Theta12 for idx 0.3893 0.04152     0.3068    0.3613   0.3896
                quant0.75 quant0.975
Theta1 for idx      0.382      0.685
Theta2 for idx      0.999      1.000
Theta3 for idx      0.997      0.999
Theta4 for idx      0.571      0.640
Theta5 for idx      0.209      0.253
Theta6 for idx      0.190      0.256
Theta7 for idx      0.069      0.131
Theta8 for idx      0.312      0.344
Theta9 for idx      0.210      0.268
Theta10 for idx     0.187      0.295
Theta11 for idx     0.269      0.325
Theta12 for idx     0.417      0.470
\end{Soutput}
\end{Schunk}

Note that the first hyperparameter in the PMCAR model is the spatial
autocorrelation, which is very close to one. All the other parameters are the
variances and correlation parameters.

In order to recover the variance matrix, the off-diagonal entries need to be
computed. Note that these depend on three parameters (i.e., correlation and
marginal variances) and that computing these entries involves multivariate
inference. As the joint posterior
distribution of these three parameters needs to be estimated sampling will
be used.

For approximate multivariate posterior inference, \pkg{INLA} can draw samples
from the (approximate) joint posterior of the hyperparameters using function
\code{inla.posterior.sample}. This sampling method is based on the internal
representation of the model, which is based on values of the ensemble
of the hyperparametes and their posterior log-densities \citep[see, for example, ][for details]{INLAbook:2020}.

The internal representation of the model stores different ensembles of values
of the hyperparameters $\{\gamma_g\}_{g=1}^G$ and associated values of the
log-posterior density. Instead of sampling with \code{inla.posterior.sample} we
will re-scale the log-posterior densities to obtain weights associated to
$\{\gamma_g\}_{g=1}^G$. By avoiding sampling we obtain short computing times
here. These weights are the posterior probabilities of the ensembles of values
of the hyperparameters and they can be used to compute posterior quantities of
interest. Furthermore, posterior estimates of transformations of the
hyperparameters can be computed as well. Note that, because the ensemble of
values is available, multivariate inference on the hyperparameters is also
possible. We will rely on this fact to estimate some of the quantities of
interest such as the posterior mean of the between-diseases variance matrix.

For the IMCAR and PMCAR models, the posterior means of the entries of the
between-diseases variance matrix are:

\begin{Schunk}
\begin{Sinput}
R> # IMCAR
R> hyper.imcar$VAR.m
\end{Sinput}
\begin{Soutput}
       [,1]   [,2]  [,3]
[1,] 0.2440 0.0811 0.130
[2,] 0.0811 0.1273 0.118
[3,] 0.1298 0.1178 0.224
\end{Soutput}
\begin{Sinput}
R> # PMCAR 
R> hyper.pmcar$VAR.m
\end{Sinput}
\begin{Soutput}
       [,1]   [,2]  [,3]
[1,] 0.2653 0.0874 0.139
[2,] 0.0874 0.1410 0.129
[3,] 0.1395 0.1286 0.255
\end{Soutput}
\end{Schunk}

For the M-model, the variance of the between diseases variance is
given by $M^{\top}M$:

\begin{Schunk}
\begin{Sinput}
R> # M-model
R> hyper.mmodel$VAR.m
\end{Sinput}
\begin{Soutput}
      [,1]  [,2]  [,3]
[1,] 0.353 0.104 0.174
[2,] 0.104 0.127 0.147
[3,] 0.174 0.147 0.226
\end{Soutput}
\end{Schunk}

Point estimates of the variance matrix are very similar for the PMCAR and the
M-model. In addition, all models seem to point to a higher correlation between
lung cancer (2nd disease in the model)  and oral cancer (3rd disease) than
cirrhosis (1st disease) with any of the two other diseases. This makes sense as
lung and oral cancer are known to be highly correlated
\citep{BotellaRocamoraetal:2015}.

Finally, the different model selection criteria computed for the models
(DIC and WAIC) can be compared to choose the best model. The values of these
criteria are shown in Table~\ref{tab:CVALIC}. In this case, both criteria
point to the M-model as the best model in this example.

\begin{table}[h!]
\centering
\begin{tabular}{c|c|c|c|c}
Model & DIC & WAIC\\
\hline
IMCAR &
7407.79 & 7396.8 \\
PMCAR &
7410.04 & 7388.57 \\
M-model &
7384.47 & 7366.15 \\
\end{tabular}
\caption{Information criteria computed for model selection.}
\label{tab:CVALIC}
\end{table}


\section{Discussion}
\label{sec:discussion}

The \pkg{INLAMSM} package builds on top of the \pkg{INLA} package and implements
a number of multivariate spatial latent effects. Hence, this package allows
an easy and simple definition of these multivariate effects to be used within
a formula term to fit multivariate spatial models to lattice data.

Implementation of new latent effects for multivariate data is straightforward
with the \code{rgeneric} latent effect included in the \pkg{INLA} package.
This only requires the specification of the latent effect as a GMRF,
which means that the mean, precision matrix and priors for the hyperparameters
need to be provided. Once the model is implemented, it is easy to include
it in the model formula to fit multivariate models with \pkg{INLA}.

We find important to mention that although all the models in \pkg{INLAMSM} have
been defined through multivariate ensembles of spatial processes, \pkg{INLAMSM}
could be also used for reproducing some multivariate spatial models defined
through univariate or multivariate conditional distributions, which induce
cross-correlations between the different spatial patterns
\citep{SainCressie:2002}.  As shown, a close connection can be drawn between
both approaches \citep{MartinezBeneito:2020} and some important classes of
conditional spatial multivariate models can be reformulated as (marginal)
multivariate models, as those reproduced in \pkg{INLAMSM}. Therefore, this
package could be used as a starting point for multivariate models formulated as
ensembles of conditional distributions

In addition, the latent models implemented in the package can be used as
templates to implement new multivariate spatial models for lattice data.  In
the future, we hope to increase the number of multivariate spatial models
in the package to include other different types of spatial cross-correlation.

In the examples provided to illustrate the use of the package we have
considered a small dataset to fit all possible models. Times required
to fit the models are short. The second example deals with a region with
a larger number of areas which shows that our package can be used
together with \pkg{INLA} to fit multivariate models.

Despite our focus on multivariate spatial models for disease mapping, it is
worth mentioning that the multivariate models implemented in the \pkg{INLAMSM}
package can be used in other contexts.  Furthermore, these multivariate spatial
models can also be used to build temporal and spatio-temporal models by using a
temporal adjacency matrix.


\section*{Acknowledgments}

This work has been supported by grants PPIC-2014-001-P and SBPLY/17/180501/000491, funded by Consejer\'ia de Educaci\'on, Cultura y Deportes (Junta de Comunidades de Castilla-La Mancha, Spain) and FEDER, grant MTM2016-77501-P, funded by Ministerio de Econom\'ia y Competitividad (Spain) and a grant to support 
research groups by the University of Castilla-La Mancha (Spain).

F. Palm\'i-Perales has been supported by a Ph.D. scholarship awarded by the University of Castilla-La Mancha (Spain).

\bibliography{MultSp-Rpack}


\end{document}